\documentclass[sn-aps,iicol]{sn-jnl}

\jyear{2022}%

\theoremstyle{thmstyleone}%
\theoremstyle{thmstyletwo}%

\theoremstyle{thmstylethree}%

\begin{document}

\title[Article Title]{\textsc{Grasian}: Towards the first demonstration of gravitational quantum states of atoms with a cryogenic hydrogen beam}

\author*[1]{\fnm{Carina} \sur{Killian}}\email{Carina.Killian@oeaw.ac.at}

\author[2]{\fnm{Zakary} \sur{Burkley}}\email{zakburkley@gmail.com}

\author[2]{\fnm{Philipp} \sur{Blumer}}\email{Philipp.Blumer@cern.ch}

\author[2]{\fnm{Paolo} \sur{Crivelli}}\email{Paolo.Crivelli@cern.ch}

\author[1]{\fnm{Fredrik P.} \sur{Gustafsson}}\email{fredrik.parnefjord.gustafsson@cern.ch}

\author[3]{\fnm{Otto} \sur{Hanski}}\email{otto.o.hanski@utu.fi}

\author[1]{\fnm{Amit} \sur{Nanda}}\email{Amit.Nanda@oeaw.ac.at}

\author[4]{\fnm{François} \sur{Nez}}\email{francois.nez@lkb.upmc.fr}

\author[5]{\fnm{Valery} \sur{Nesvizhevsky}}\email{nesvizh@ill.fr}
 
\author[4]{\fnm{Serge} \sur{Reynaud}}\email{serge.reynaud@lkb.upmc.fr}

\author[2,5]{\fnm{Katharina} \sur{Schreiner}}\email{schreink@student.ethz.ch}

\author[1]{\fnm{Martin} \sur{Simon}}\email{Martin.Simon@oeaw.ac.at}

\author[3]{\fnm{Sergey} \sur{Vasiliev}}\email{servas@utu.fi}

\author[1]{\fnm{Eberhard} \sur{Widmann}}\email{Eberhard.Widmann@oeaw.ac.at}

\author[4]{\fnm{Pauline} \sur{Yzombard}}\email{pauline.yzombard@lkb.upmc.fr}

\affil[1]{\orgdiv{Stefan Meyer Institute for Subatomic Physics}, \orgname{Austrian Academy of Sciences}, \orgaddress{\street{Kegelgasse 27}, \city{Vienna}, \postcode{1030}, \country{Austria}}}

\affil[2]{\orgdiv{Institute for Particle Physics and Astrophysics}, \orgname{\textsc{eth}, Zurich}, \orgaddress{\city{Zurich}, \postcode{8093}, \country{Switzerland}}}

\affil[3]{\orgdiv{Department of Physics and Astronomy}, \orgname{University of Turku}, \orgaddress{ \city{Turku}, \postcode{20014}, \country{Finland}}}

\affil[4]{\orgdiv{Laboratoire Kastler Brossel}, \orgname{Sorbonne Université, CNRS, ENS-PSL Université, Collège de France}, \orgaddress{\city{Paris}, \postcode{75252},  \country{France}}}

\affil[5]{\orgdiv{Institut Max von Laue - Paul Langevin}, \orgaddress{\street{71 avenue des Martyrs}, \city{Grenoble}, \postcode{38042}, \country{France}}}


\abstract{
At very low energies, a light neutral particle above a horizontal surface can experience quantum reflection. The quantum reflection holds the particle against gravity and leads to gravitational quantum states (\textsc{\textsc{gqs}}).
So far, \textsc{gqs} were only observed with neutrons as pioneered by Nesvizhevsky and his collaborators at \textsc{ill}.
However, the existence of \textsc{gqs} is predicted also for atoms. 
\\
The \textsc{Grasian} collaboration pursues the first observation and studies of \textsc{\textsc{gqs}} of atomic hydrogen. 
We propose to use atoms in order to exploit the fact that orders of magnitude larger fluxes compared to those of neutrons are available. 
Moreover, recently the \textit{q}-\textsc{Bounce} collaboration, performing \textsc{gqs} spectroscopy with neutrons, reported a discrepancy between theoretical calculations and experiment which deserves further investigations. 
For this purpose, we set up a cryogenic hydrogen beam at \SI{6}{\K}. We report on our preliminary results, characterizing the hydrogen beam with pulsed laser ionization diagnostics at  \SI{243}{\nano\m}.
}




\maketitle
\section{Introduction}\label{sec1}

Quantum bouncers were first predicted in 1928 \cite{Bre:1928pr}. Nearly 75 years later, this phenomenon was demonstrated through the observation of neutron ($n$)  gravitational quantum states (\textsc{gqs}) \cite{Lus:1978jl,Nes:2000nima,Nes:2002Nat,Nes:2003prd,Nes:2003prdbis,Nes:2005epjc,Wes:2007epjc}.
Confined by the gravitational- and the mirror potential, the $n$ are settled in gravitationally bound quantum states.

Studies of $n$ \textsc{gqs} have a broad impact on fundamental and applied physics. 
They serve as a unique method to study the interaction of a particle in a quantum state with a gravitational field. 
For example, paired with more recent measurements of $n$ whispering gallery states (\textsc{wgs}) - quantum states trapped by the centrifugal- and the mirror potential \cite{Nes:2010NatPhys}, they result in the first direct demonstration of the validity of the weak equivalence principle for a particle in a pure quantum state.

The observation of \textsc{gqs} initiated active analysis of the pecularities of this phenomenon \cite{Rob:2004pr,Ber:2005jmc,Mat:2006pra,ber:2006plb,Rom:2007prl,Del:2009prl,Gar:2012pr,Bel:2014pr} and their application to the search for new physics, such as the searches for extra fundamental short-range interactions \cite{Abe:2003lnp,Fra:2004plb,Bra:2006prd,Bui:2007cqg,Bae:2007prd,shortrangef,Bra:2011prl}, verification of the weak equivalence principle in the quantum regime \cite{Ber:2003CQG,Kie:2005AP,Kaj:2010APB}, extensions of quantum mechanics \cite{Ber:2005PRD,Sah:2007EPJC}, extensions of gravity and space theories \cite{Noz:2010EPL,Ped:2011JHEP,Kob:2011PRD,Cha:2011PLB} or tests of Lorentz invariance \cite{Mar:2018PRD,Esc:2019PRD,Iva:2019PLB}.

New fundamental short-range interactions are predicted in extensions of the Standard Model such as grand unified, supersymmetric and string theories that could alter the weak gravitational potential. They also appear in certain models explaining dark matter and dark energy \cite{shortrangef}. 
Additionally, studies on \textsc{gqs} can provide extremely sensitive measurements of the mirror’s surface potential and shape, which is of high interest to the surface physics community.

Spectroscopy and interferometry methods of observation of \textsc{gqs} of $n$ have been analyzed theoretically and implemented experimentally over the previous two decades \cite{Nes:2006TQGR,Nes:2010PU,Jen:2011NP,Jen:2014PRL,Ich:2014PRL}. 
However, the existence of \textsc{gqs} is predicted also for atoms and antiatoms \cite{Vor:2011PRA, WOS:000395981900005, WOS:000312994700008, WOS:000225326300020, WOS:000186422700011, WOS:A1993LW23100029, WOS:A1989AU53000009, WOS:A1972N687600001, WOS:000222126500001}. 
Those are expected to have essentially identical properties for particles of almost equal mass such as $n$, atomic hydrogen ($H$) or even antihydrogen ($\bar{H}$). 

A major constraint to improve the precision of the current measurements of \textsc{gqs} of $n$ is the limited density of ultracold $n$ (\textsc{ucn}s). It looks natural to exploit the much higher fluxes available for atoms, namely the high densities of existing $H$-beams \cite{CooperHBeam2020}.
However, all the projects concerning the use and study of \textsc{gqs} of atoms are currently based only on theoretical estimations since those, in contrast to $n$, have never been observed experimentally. Only a direct experiment can prove the existence of \textsc{gqs} of atoms, evaluate the systematic and statistical uncertainties of such experiments and develop the experimental techniques needed for more precise measurements in the future. 

In section \ref{sec:gqs_theory}, a theoretical derivation of \textsc{gqs} is given.
The method used earlier for the observation of $n$ \textsc{gqs} and the planned implementation of a measurement with $H$ is presented in \ref{sec:gqs_methods}. A detailed description of the \textsc{Grasian} experimental setup and the recent measurements is given in \ref{sec:gqs_setup}.

\section{Theoretical framework}
\label{sec:gqs_theory}
A sufficiently slow particle trapped by the gravitational field on one side and a horizontal reflective surface ("mirror") on the other side settles in \textsc{gqs}.
The particle's wave function $\psi (z)$ in the Earth's gravitational field above a mirror is governed by the Schrodinger equation $\frac{\hbar ^2}{2m} \frac{d^2 \psi (z)}{dz^2} + (E-mgz) \psi (z) =0$, where $\hbar$ is the reduced Planck constant, $m$ is the particles mass, $z$ is the height, $E$ is the energy of the vertical motion of the particle, and $g$ is the acceleration in the Earth's gravitational field. The only constant related to the particle's identity is it's mass, which is nearly exactly the same for $n$, $H$ or $\bar{H}$. 
For simplicity, $n$ over an ideal mirror will be considered in the following derivation.

An ideal horizontal mirror at the height $z=0$ can be approximated as an infinitely high and abrupt potential step. This approximation is justified by the characteristic values of energies and lengths in our problem. The energy of neutrons in low quantum states $\sim$\SI{e-12}{\eV}, is much smaller than the optical potential of the mirror material $\sim$\SI{e-7}{\eV} \cite{golub1991ultra}, and the characteristic range of increase in the optical potential for a polished mirror $\sim$\SI{e-9}{\m}, is much smaller than the wavelength of neutrons in low quantum states $\sim$\SI{e-5}{\m}. Such an infinitely high and abrupt optical potential corresponds to the zero boundary condition for the wave function, $\psi (z=0)=0$. 

A solution of the Schrodinger equation can be written in terms of the Airy function Ai, $\psi (z)=C \text{Ai}(z/z_0)$, where $z_0=[\hbar^2/(2m^2g)]^{1/3}=\,$\SI{5.87}{\micro\m} is the characteristic length scale of the problem and $C$ is a normalization constant. The Airy function zeros $\lambda_k$ define the quantum state energies $E_k=mgz_0\lambda_k$, where $\varepsilon_0=mgz_0=\,$\SI{0.602}{\pico\eV} is the characteristic energy of the problem and $f_0=\varepsilon_0/(2\pi \hbar)=\,$\SI{145}{\Hz} is its characteristic frequency. 

The five lowest zeros of the Airy function Ai are $\lambda_k=\{2.34, 4.09, 5.52, 6.79, 7.94...\}$. The eigenfunctions of the quantum states are $\psi_k(\xi(z))\sim C_k \text{Ai}(\xi_k(z))$, where $\xi_k(z)=z/z_0-\lambda_k$, and $C_k$ are normalization constants.

\begin{figure}[h!]
\centering
\includegraphics[width=0.4\textwidth]{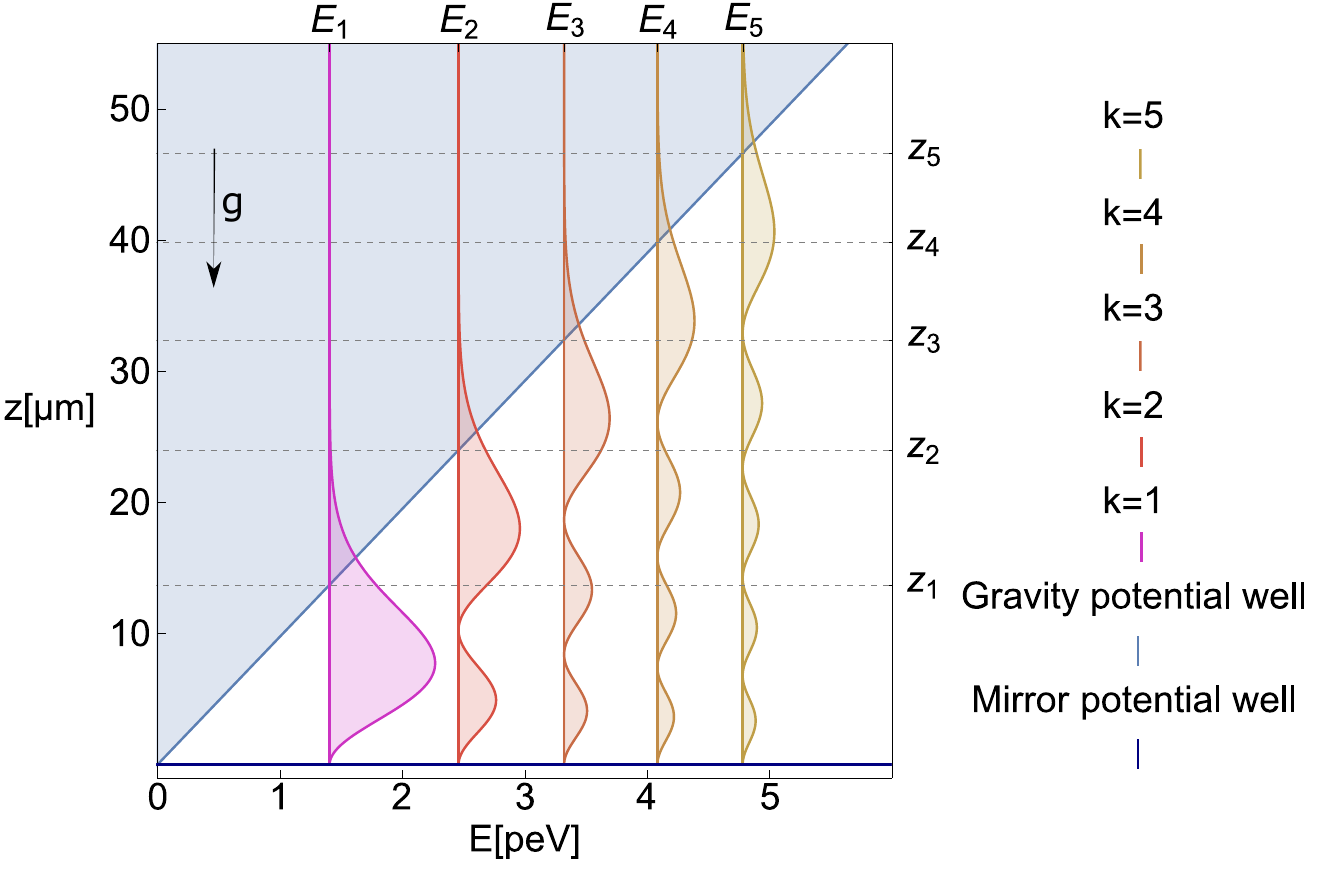}
\caption{Squared modules of the neutron wavefunctions $\abs{\psi_k(z)}^2$ as a function of the height $z$ for the five lowest quantum states; they correspond to the probabilities of observing neutrons at a height $z$.}
\label{fig:gqs1}
\end{figure}

The energy eigenvalues $E_k$ depend only on $m$, $g$ and $\hbar$, and are independent of the ideal mirror properties. Within the classical description, a neutron with the energy $E_k$ can rise in the gravitational field up to the height $z_k=E_k/{mg}$. In quantum mechanics, the probability of observing a neutron with the energy $E_k$ in the $k^{th}$ quantum state at a height $z$ is equal to the squared modulus of its wave function (see  Fig. \ref{fig:gqs1}). 

\section{Measurement method}
\label{sec:gqs_methods}

\subsection{\textsc{gqs} measurement with $n$}
\label{sec:ucn_methods}
\begin{figure}[h]
\centering
\includegraphics[width=0.45\textwidth]{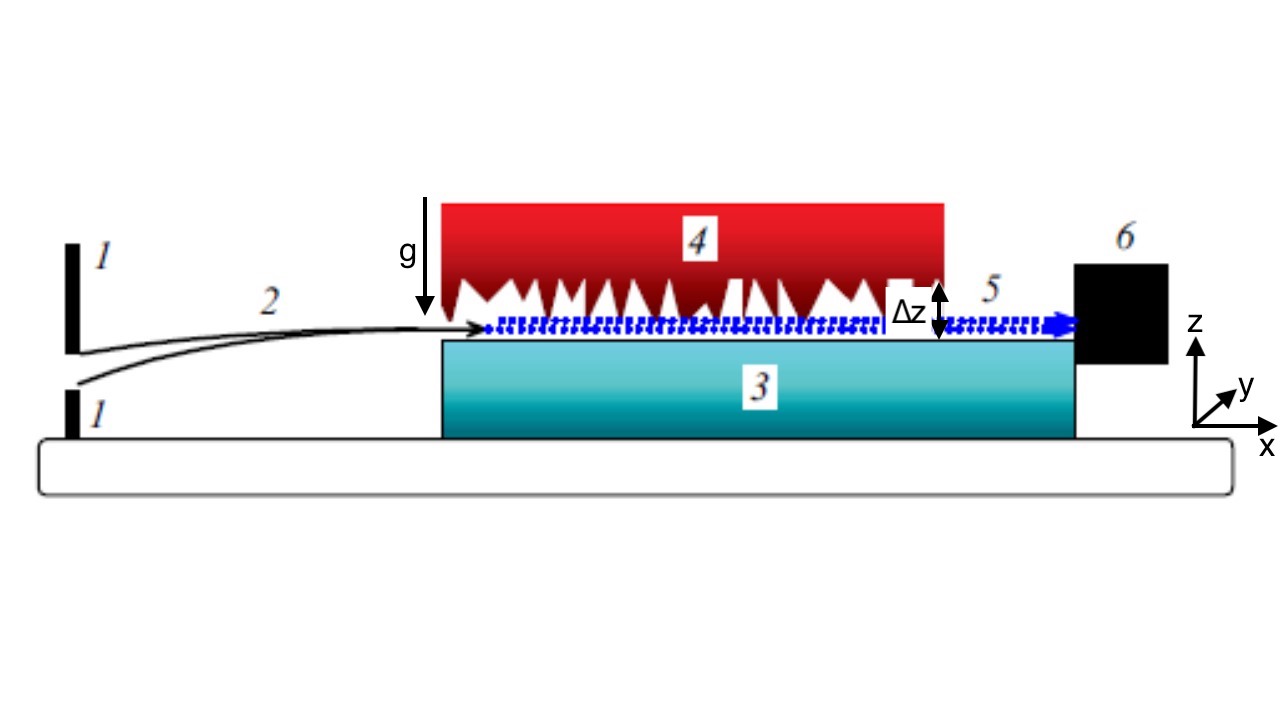}
\caption{Schematic of the experimental setup in the flow through mode. 1 are the bottom and top entrance collimator plates, arrows 2 correspond to neutron classical trajectories between the collimator and the entrance to the slit between mirror 3 and scatterer 4. Dotted horizontal arrows 5 illustrate neutron quantum motion above the mirror. 6 is the neutron detector. The height of the slit between the mirror and the scatterer can be varied and precisely measured.} \label{fig: Transmission}
\end{figure}

In this section, the methods developed for the first observation of \textsc{gqs} of $n$ at \textsc{ill} will be described.
The experimental installation is a one component gravitational \textsc{ucn} spectrometer with a high energy and spatial resolution \cite{Nes:2000nima}. The principle of its operation, illustrated in  Fig. \ref{fig: Transmission}, is the measurement of the $n$ flux through a slit between the mirror on bottom and the flat scatterer on top as a function of the slit height $\Delta z$ which can be changed and precisely measured. 

The scatterer surface is smooth on a large scale but rough on the \si{\micro\m} scale. The roughness amplitude is about a few \si{\micro\m}, and is comparable to the characteristic scale $z_0$ of the problem. 
The scatterer's surface reflects $n$ which reach it non specularly, mixing the vertical and horizontal velocity components of the $n$. Because the $n$ horizontal velocity components are much larger than their vertical velocity components, such mixing causes numerous collisions of the $n$ with the scatterer, thus resulting in a prompt loss of those $n$.

The length of the bottom mirror is chosen based on the energy time uncertainty relation $\Delta\tau \Delta E \ge \hbar/2$ . The observation of the $k^{th}$ quantum state is possible if the difference between the Eigenenergies of state $k+1$ and state $k$, $\Delta E_{k+1,k}$ is bigger than the width of the $k^{th}$ quantum state, $\delta E_k$: $\Delta E_{k+1,k}>\delta E_k$. As the state number $k$ increases, $\Delta E_{k+1,k}\sim k^{-1/3}$ decreases until the levels pass into the classical continuum. Evidently, measurements of low quantum states are easier and more convenient. $\delta E_k$ is defined by the time of flight of $n$ above the mirror. Therefore, the mirror length is determined by the time interval needed to observe a $n$ in a \textsc{gqs}: $\Delta\tau\sim\SI{0.5}{\milli\s}$. It follows, that the mirror length should be  $L\sim\SI{10}{\centi\m}$ for low states and for $n$ velocities $v_{\rm{hor}}\sim5-$\SI{10}{\m\per\s}.

The vertical scale in the problem is defined by the relation between momentum $p$, velocity $v$ and wavelength $\lambda$: $p=mv=h/\lambda$ and the momentum position uncertainty relation $\Delta p \Delta z  \ge \hbar/2$.
The smaller the $n$ vertical velocity component, the larger the $n$ wavelength associated with this velocity component. But the classical height up to which a $n$ can rise in the gravitational field cannot be smaller than the quantum mechanical uncertainty of its vertical coordinate, i.e., the $n$ wavelength. This relation determines the lowest bound state of $n$ in the Earth's gravitational field. The height uncertainty is then $\Delta z\sim z_0$, and the vertical velocity uncertainty is $\Delta v_z\sim v_0=\sqrt{2\varepsilon_0 /m}=$\SI{1.07}{\centi\m\per\s}, the characteristic velocity in the problem.

\begin{figure}[h]
\centering
\includegraphics[width=0.4\textwidth]{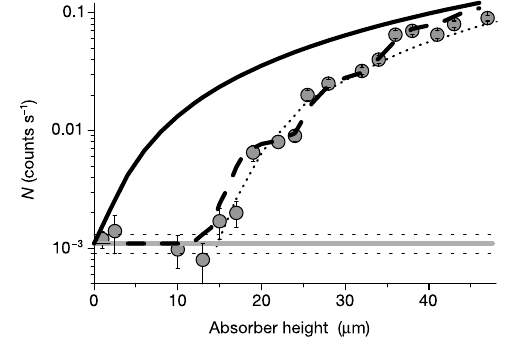}
\caption{The data points correspond to the measured $n$ flux through the slit between mirror and absorber versus the slit width at low width values. The dashed curve is a fit using the quantum mechanical calculation, where all level populations and the height resolution are extracted from the experimental data. The solid curve is the full classical treatment. The dotted line is a truncated fit in which it is assumed that only the lowest quantum state - which leads to the first step - exists. Fig. taken from \cite{Nes:2002Nat}.}
\label{fig: NeutronData}
\end{figure}

The method used in the first observation of $n$ \textsc{gqs} \cite{Nes:2002Nat} consisted in measuring $n$ transmission through the narrow slit $\Delta z$ between the horizontal mirror and the scatterer above it. 
If $\Delta z\gg z_k$, neutrons in the $k^{th}$ quantum state pass through the slit with no significant loss. But as $\Delta z$ decreases, the neutron wave function $\psi_k(z)$ starts penetrating the scatterer, and the $n$ loss probability increases. If $\Delta z\leq z_k$, the slit is practically nontransparent to neutrons in the $k^{th}$ quantum state. In an "ideal" experiment with an infinitely high energy resolution, the $n$ flux $N_{\rm{QM}}(\Delta z)$ through the slit would sharply change at the height $\Delta z\sim z_k$. 
In reality, the idealized step like dependence is smoothed due to two factors: the spectrometer experimental resolution and the smooth shape of $n$ wave functions. The latter is due to the tunneling of $n$ through the gravitational barrier separating the classically allowed heights and the scatterer height.
An example of the experimental data is shown in  Fig. \ref{fig: NeutronData}.

\subsection{\textsc{gqs} measurement with $H$}
\label{sec:qgsH}
The method, developed for the observation of \textsc{gqs} of $n$, will be used to demonstrate \textsc{gqs} of $H$. In the following section, the feasibility of such an experiment will be analysed and the characteristic parameters of the two experiments will be compared.

The observation time is a key parameter.
The time of observation is defined by the mean particle velocity and the mirror length. For the characteristic velocity of $n$, $v_n\sim\,$\SI{10}{\m\per\s} and the mirror length $L_n=\,$\SI{10}{\centi\m}, the observation time was $\tau_n\sim\,$\SI{10}{\milli\s}. $\tau_n / \tau_0\sim 20$, i.e. the observation time is much larger than the formation time ($\tau_n\gg\tau_0$) and the \textsc{gqs} were well resolved. In order to provide the same conditions for the resolution of \textsc{gqs}, the time of observation of $H$ has to be the same $\tau_H\sim\tau_n\sim\,$\SI{10}{\milli\s}. For the planned \textsc{gqs} mirror length $L_H\sim \,$\SI{30}{\centi\m}, the mean velocity of $H$ has to be $v_H\sim L_H/\SI{10}{\milli\s}\sim\SI{30}{\m\per\s}$. All $H$ with significantly higher velocities do not settle in \textsc{gqs}, they only increase the background and have to be eliminated. 
Velocities of up to $v_H\sim \SI{100}{\m\per\s}$ can still be tolerated, at the cost of a worse energy resolution of the experiment. This is still acceptable for the first observation of \textsc{gqs} of $H$. However, low velocities, good control over the velocity selection and sufficient background suppression is the key condition for the observation of \textsc{gqs} of $H$.

A major difference in the behavior of $n$ and $H$, is the mechanism of their interaction with the rough surface of the scatterer. Scattered $n$ are lost in the bulk of the mirror or scatterer after several reflections from their surfaces. Scattered $H$ atoms have higher chances to leak through the slit and produce background. They escape from the slit into a broad angular distribution, in contrast to $H$, which pass the slit setteld in a \textsc{gqs}. $H$ atoms escaping to larger angles have to be eliminated to decrease the background. Therefore, it is very important to implement a proper background suppression.

The expected count rate of $H$ is much higher than that of $n$. A simple comparison of the "brightness" of the $n$ source at \textsc{ill} and the $H$ source at the \textsc{Grasian} experiment at \textsc{eth} Zurich is the following. The total number of the particles produced at \textsc{ill} is $\sim\SI{e19}{\per\s}$, while it is $\sim\SI{e17}{\per\s}$ at \textsc{eth}. A characteristic temperature of $n$ spectrum at \textsc{ill} is $\sim \SI{40}{\K}$, while it is $\sim \SI{6}{\K}$ at \textsc{eth}. An effective surface area of the source at \textsc{ill} is $\sim \SI{e5}{\centi\m\squared}$, while it is $\sim \SI{1}{\centi\m\squared}$ at \textsc{eth}. These values result in a $\sim 10^4$ higher flux of particles at \textsc{eth}. This estimation does not account for the velocity and spatial distributions of the particles at the two sources, the detection efficiencies, the transport losses, the experiment geometries etc. but gives a fair overall gain factor and the conclusion that $H$ fluxes are much higher than those available at the best existing \textsc{ucn} sources. A more detailed calculation provides the same estimate of the count rate for this setup: $\sim 10^3$ $H\si{\per\s}$ for the lowest \textsc{gqs}, over four orders of magnitude larger than measured with $n$. 
With these increased count rates, and the minimal background with the $H$ detection method based on the photoionization described in section \ref{sec:laser}, one can measure the height distribution covering the first four \textsc{gqs} with a \SI{1}{\micro\m} resolution and 5$\%$ accuracy in a matter of hours (compared to several days with the original $n$ experiment). These robust statistics will give fast feedback, enabling us to solve experimental challenges quickly and efficiently.

\section{\textsc{Grasian} cryogenic $H$-beam}
\label{sec:gqs_setup}

In this section the \textsc{Grasian} cryogenic $H$-beam, located at \textsc{eth} Zurich, will be described. It was originally developed in the context of the ASACUSA \cite{Diermaier}, MuMASS \cite{2018-MuMASS,PhysRevLett.128.011802} and GBAR \cite{Perez:2015zya} experiments to test and commission detectors, microwave- and laser-setups.
It will be used for our attempt to demonstrate \textsc{gqs} of atoms for the first time.

The \textsc{eth} $H$-beam consists of an $H$ source, a cryogenic chamber, a beamline and a detection chamber. 
The \textsc{gqs} chamber, which will contain the mirror-absorber system described in section \ref{sec:gqs_methods}, will be installed in the near future.

\begin{figure}[h]
\centering
\includegraphics[width=0.5\textwidth]{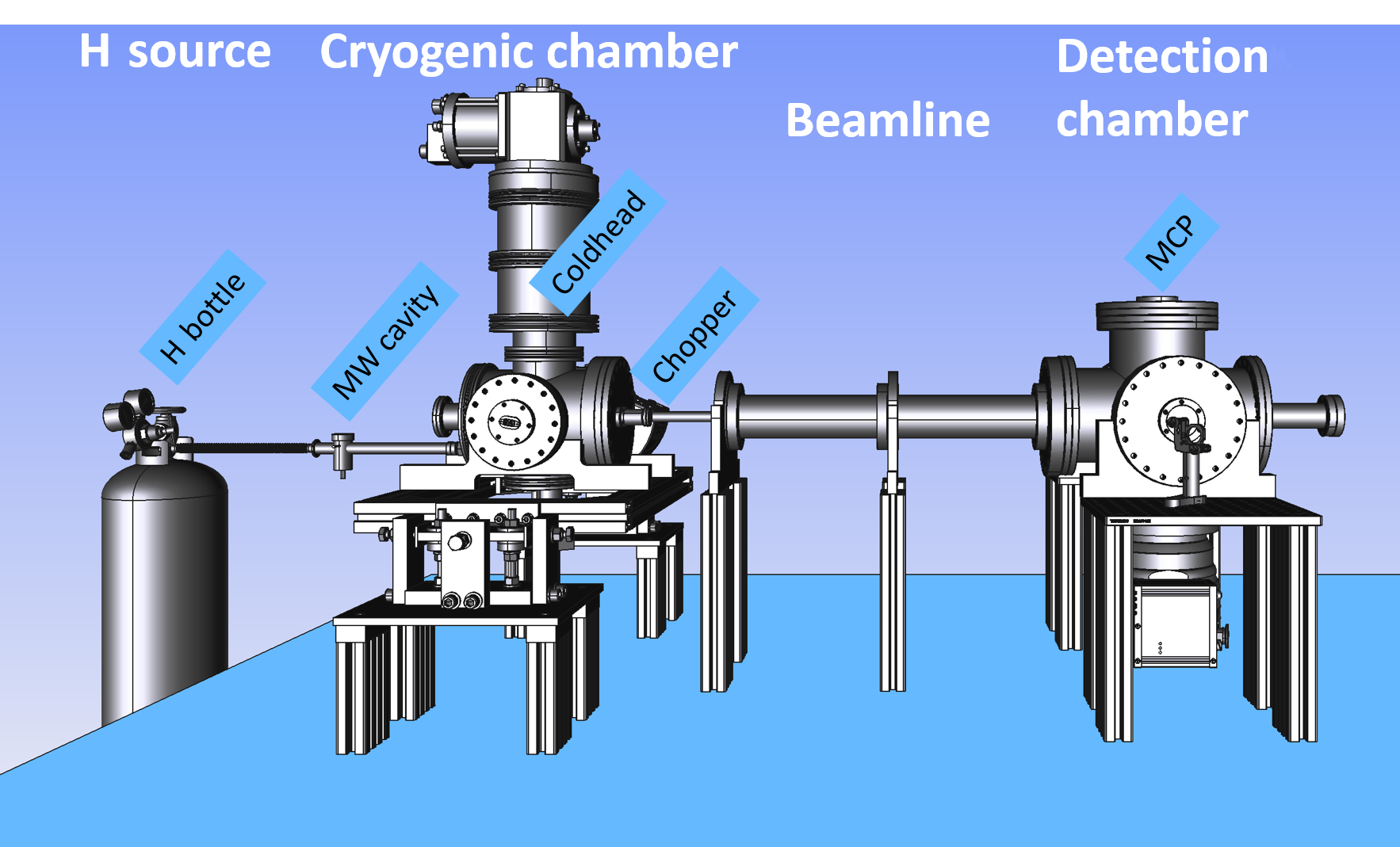}
\caption{Rendering of the \textsc{eth} cryogenic $H$-beam setup.} 
\label{fig: setup}
\end{figure}
In the $H$ source, molecular hydrogen gas ($H_2$) is injected into a microwave cavity with a flow of $\sim\SI{0.8}{\milli\liter\per\min}$. Within the cavity, microwaves with 20-\SI{30}{\W} at \SI{2400}{\mega\Hz}, dissociate the $H_2$ into atomic $H$.
The cracking efficiency was measured to be $\sim 80\%$, resulting in an $H$ flux of more than \SI{e17}{atoms\per\s}. 

The generated atomic $H$ is directed into the cryogenic chamber through a glass-teflon tube. The tube is inserted into a bent nozzle, directly connected with a coldhead to thermalize the atoms to $\sim\SI{6}{\K}$ with minimal recombination. 
After the cryogenic chamber, the beam of atomic $H$ is divided into bunches of $\sim$\SI{e14}{atoms\per\s} by a rotating chopper wheel. The $H$ bunches flow through a system of velocity selecting apertures into the detection chamber at the end of the beamline.

A \SI{243}{\nano\m} ultraviolet (UV) laser is directed into the detection chamber in a perpendicular direction with respect to the atomic beam and is retro reflected by a mirror on the other side.
This setup creates two counter propagating beams, as required to excite the hydrogen atoms from the 1S to the 2S state via a two photon transition. A third photon from the same laser induces photoionization, the resulting protons can be detected by a microchannel plate (MCP).

\subsection{The laser system for hydrogen detection}
\label{sec:laser}
The $H$ 1S-2S transition is dipole forbidden, hence it is only possible to drive it with two photons. The energy of this transition is split between the two photons involved, meaning that the wavelength of the laser driving this transition has to be $\sim\SI{243}{\nano\m}$.
As a dipole forbidden transition is very unlikely, a high photon density and hence a high intensity laser is needed for efficient excitation and subsequent ionization of $H$ atoms.
The most efficient realization is a pulsed laser. Such a laser system has been set up at \textsc{eth} Zurich. 
It will be described in the following section.

\begin{figure}[h]
\centering
\includegraphics[width=0.45\textwidth]{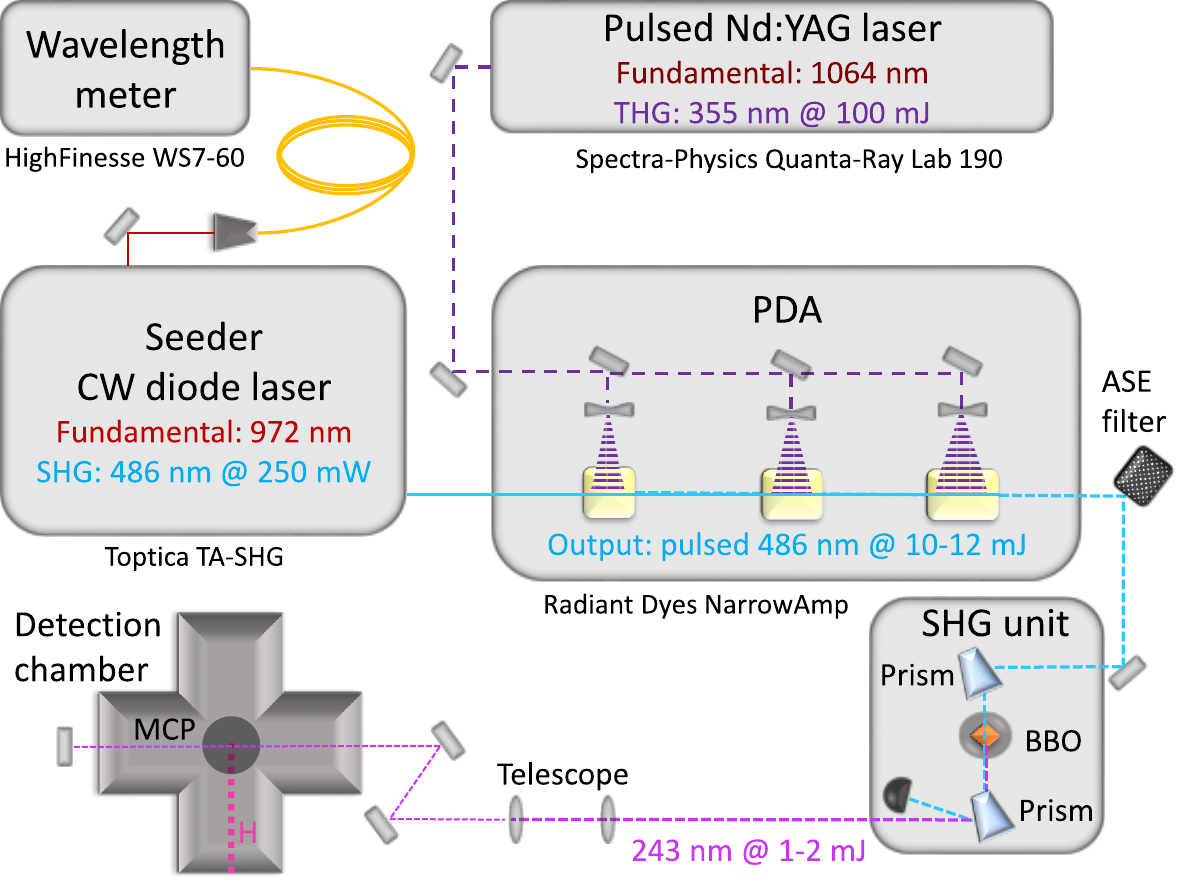}
\caption{Schematic of the laser system. The \textsc{pda} is pumped with a pulsed Nd:YAG laser and seeded with a CW diode laser. The generated \SI{486}{\nano\m} pulsed beam is frequency doubled in the \textsc{shg} unit and afterwards sent into the detection chamber to ionize $H$. The wavelength of the fundamental of the seeder laser, is determined and controlled by a wavelength meter.}
\label{fig: laser_setup}
\end{figure}

A schematic of the laser system is shown in Fig. \ref{fig: laser_setup}. It consists of two lasers - a continuous wave\footnote{Toptica TA-SHG pro} (CW) and a pulsed laser\footnote{Spectra-Physics Quanta-Ray Lab 190}, a pulsed dye amplifier\footnote{Radiant Dyes NarrowAmp} (\textsc{pda}) and a second harmonic generation (\textsc{shg}) unit.
The fundamental of the CW laser has a wavelength of \SI{972}{\nano\m}. An internal \textsc{shg} cavity creates the output CW laser beam with a wavelength of \SI{486}{\nano\m} and an average power of $\sim$\SI{250}{\milli\W}. With exactly double of the desired wavelength for the two photon 1S-2S transition, the CW laser acts as the seeder laser.   
The pulsed laser is the \SI{355}{\nano\m} third harmonic of a \SI{10}{\Hz} pulsed Nd:YAG laser. In the Q-switch mode, the pulses are \SI{10}{\nano\s} long and have a pulse energy of $\sim$\SI{100}{\milli\J}.

Within the \textsc{pda}, three quartz cuvettes are circulated with the dye, Coumarin 102, dissolved in ethanol. The absorption and fluorescence of Coumarin 102 fit the application well: when pumped with \SI{355}{\nano\m} light, the emission is centered around \SI{473}{\nano\m}.
The pulsed laser is split up and focused onto the three cuvettes, and pumps the Coumarin molecules. The CW beam seeds the \textsc{pda} by passing through the three cuvettes overlapping with 
the pumped dye molecules. This stimulates the emission of \SI{486}{\nano\m} photons with every pulse that pumps the dye. After three cuvettes, $\sim\,$10-\SI{12}{\milli\J} of \SI{486}{\nano\m} pulsed laser light is generated. Like the pulsed pump laser, it runs at \SI{10}{\Hz} with a pulse length of $\sim\SI{10}{\nano\s}$. 

\begin{figure}[h]
\centering
\includegraphics[width=0.45\textwidth]{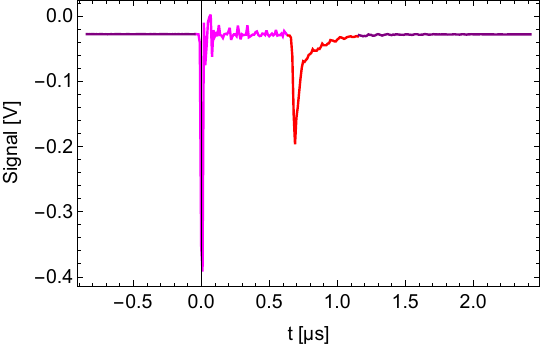}
\caption{Waveform captured by the oscilloscope showing the $H^+$ signal.
The purple region determines the offset in the signal strength evaluation. 
The magenta curve corresponds to the signal induced by the UV light in the detection chamber. The red curve corresponds to the $H^+$ induced voltage change in the expected time-of-flight window between \SI{640}{\nano\s} to \SI{1.15}{\micro\s}.}
\label{fig: ROI of Hydrogen}
\end{figure}

In the \textsc{shg} unit, the output of the \textsc{pda} is frequency doubled with a barium borate (\textsc{bbo}) crystal to generate $\sim$1-\SI{2}{\milli\J} of pulsed UV radiation at \SI{243}{\nano\m}. 
The \SI{243}{\nano\m} beam is then sent into the detection chamber.
A mirror is mounted behind the photo ionization region to produce counter-propagating beams for the doppler free two photon excitation. 
The \SI{243}{nm} photons efficiently ionize the $H$ and the $H^+$ ions are detected by an MCP. The MCP creates a voltage signal, which is read out by an oscilloscope.
In Fig. \ref{fig: ROI of Hydrogen}, such a waveform of $H^+$ signal is shown. 

As indicated in Fig. \ref{fig: laser_setup}, the frequency of the fundamental of the seeder laser is determined and controlled by a wavelength meter\footnote{HighFinesse WS7-60}. While the wavelength meter is reliable for relative measurements, the absolute values are shifted by $\sim\SI{230}{\mega\Hz}$, due to outdated calibration.
The CW fundamental frequency corresponds to $1/8$ of the 1S-2S transition frequency, due to two \textsc{shg} processes and the two photon excitation.

\begin{figure}[h]
\centering
\includegraphics[width=0.45\textwidth]{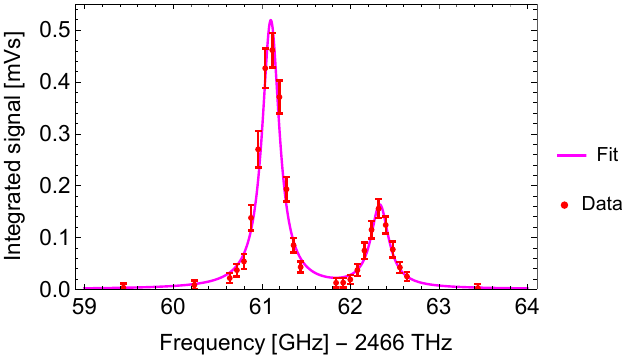}
\caption{Laser frequency sweep - Observation of the 1S and 2S HFS of $H$. 
The frequency values on the abscissa correspond to the measured frequency of the seeder laser fundamental, shifted by $\SI{230}{\mega\Hz}$ (outdated wavelength meter calibration) and multiplied by a factor of 8 (two SHG processes, two photon excitation). This was done to to match the absolute literature values \cite{Parthey:2011lfa}. }
\label{fig: HFS}
\end{figure}

Scanning the laser frequency around the resonance, shows that we are capable to resolve the hyperfine splitting (HFS) of $H$. 
The two peaks, shown in Fig. \ref{fig: HFS}, correspond to the difference of the HFS of the 1S and the 2S state. The measured value is $\Delta\nu_{\rm{meas.}}=\SI{1.23 \pm 0.02}{\giga\Hz}$, which agrees with the literature value $\Delta\nu_{\rm{lit.}}=\SI{1.24}{\giga\Hz}$ (which can be calculated from \cite{Parthey:2011lfa}) within $1\sigma$. This proves, that we detect $H$ atoms, and not any other potential pollutant in the detection chamber.

A scan of the laser pulse intensity dependency of the signal resulted in the same conclusion, showing the expected behavior.
The intensity $I$ is determined by the pulse energy $E$ and the beam waist $\omega_0\sim\SI{0.75}{\milli\m}$: $I=E/(\omega_0^2\pi)$.

\begin{figure}[h]
\centering
\includegraphics[width=0.45\textwidth]{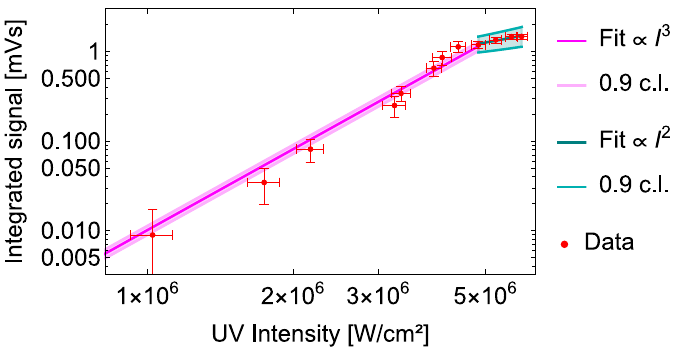}
\caption{Laser intensity scan - Observation of the $I^3$ and $I^2$ dependency of $H$ ionization.}
\label{fig:energy}
\end{figure}

The data can be taken from Fig. \ref{fig:energy}. 

The signal shows an $I^3$ dependence as expected for a three photon process. The two photon 1S-2S excitation is $I^2$ dependent and the ionization of the 2S state adds another $I$ dependency to the overall process.
At $\sim\SI{E7}{\W\per\centi\m\squared}$, the 2S ionization process starts to saturate, as the ionization rate $\Gamma_i=(I/{h\nu})\sigma$, where $\sigma$ is the 2S $H$ ionization cross section and $h\nu$ the photon energy, reaches $1/\tau_{\rm{2S}}$, where $\tau_{\rm{2S}}$ is the lifetime of the 2S state. From here, the overall process follows an $I^2$ dependency. 
At $\sim \SI{3E7}{\W\per\centi\m\squared}$, also the 1S-2S excitation process will start to saturate and the overall process becomes $I$ independent \cite{Downey:89}.

It would be ideal to run the $H$ detection on saturation, because the ionization process would become independent of energy- or frequency instabilities of the laser.
In order to reach the point of saturation, the laser beam size has to be decreased. Different combinations of convex and concave lenses were already used to compress the beam.
But, the mirrors did not withstand the increased intensity for long and were damaged. It seemed like the point of saturation overlapped with the damage threshold of the UV mirrors, which were used at that point. We replaced the mirrors and will hopefully be able to increase the intensity until saturation is reached.


\subsection{Hydrogen beam characterization and rate estimation}
To characterise the velocity distribution of the $H$-beam, a time of flight (ToF) measurement was performed. The delay between the opening of the chopper and the firing of the laser was varied while the $H$ count rate was measured.

The expected signal $S(t)$ is a convolution of the chopper kernel $h(t)$ and the atomic ToF distribution $P_t(t)$ (assumed to follow a Maxwellian distribution) and is given by
\begin{align}
S(t) &= h(t) * P_t(t)\, , \\
P_v(t) & \propto v^3 \exp{\left(-\frac{mv^2}{2kT}\right)}\, , \\
P_t(t) &= P_v\left(\frac{\Delta x}{t}\right) \frac{\Delta x}{t^2}\, ,
\end{align}
where $\Delta x$ is the distance between chopper and detection region, $m$ is the $H$ mass, $k$ is the Boltzmann constant and $T$ the temperature.

\begin{figure}[h]
\centering
\includegraphics[width=0.45\textwidth]{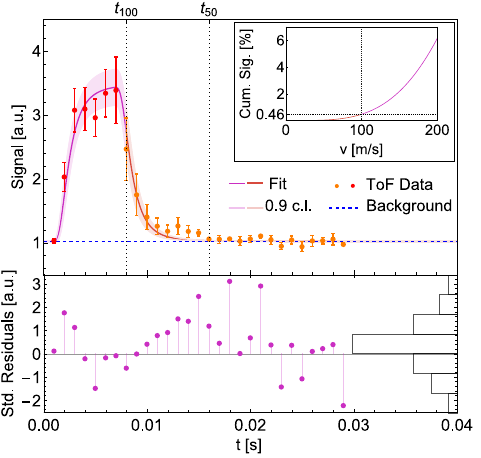}
\caption{Upper figure: ToF data and fit. The fit results in a temperature of $T=\SI{6.07+-0.74}{\K}$. $t_{100}$ and $t_{50}$ correspond to the ToFs of $H$ with velocities of \SI{100}{\m\per\s} and \SI{50}{\m\per\s}, respectively. In the small figure, the corresponding cumulative velocity distribution for $v\in[0,200]\,\si{\m\per\s}$ is shown.
Lower figure: Standardized residuals of the fit and corresponding histogram.}
\label{fig: delay}
\end{figure}

The data taken in 2021 and a fit are shown in Fig. \ref{fig: delay}. The fit resulted in a temperature of $T=\SI{6.07+-0.74}{\K}$, meaning, that the $H$ gas thermalizes well with the cryogenic nozzle at \SI{6}{\K}.
The measurement shows that the maximum of the atom flux appears after around \SI{5}{\milli\s} delay, relating to an atomic velocity of \SI{250}{\m\per\s} with a significant fraction of atoms below \SI{100}{\m\per\s}.

As mentioned in section \ref{sec:qgsH}, $H$ velocities of up to \SI{100}{\m\per\s} can be tolerated. For the upcoming \textsc{gqs} measurement, a certain velocity interval will be selected by setting the delay between chopper opening and firing of the laser to a certain value. The width of  this interval is determined by the duration of the chopper opening $t_{\rm{open}}\sim\SI{6.1}{\milli\s}$.
With the current velocity distribution of the $H$ beam, it makes sense, to set the upper bound of the velocity interval to $v_{\rm{max}}=\SI{100}{\m\per\s}$ which leads to a lower bound of $v_{\rm{min}}=\Delta x / (t_{100}+t_{\rm{open}}) = \SI{62}{\m\per\s}$. The mean velocity of this interval is $\bar{v}=\SI{81}{\m\per\s}$ with a corresponding ToF of $t_{\bar{v}}=\SI{12.3}{\milli\s}$.

With the fit result, it is possible to estimate the rate of $H$ atoms passing through the future \textsc{gqs} region. It is composed of the $H$ input rate $R_{\rm{in}}\sim\SI{E17}{\per\s}$, the chopper duty cycle $d_c\sim0.061$, the form of the distribution after the chopper, assuming a cone like distribution with an opening angle of  $\theta\sim\frac{3}{4}\pi$, the cross sectional area of the \textsc{gqs} region $A\sim\SI{0.5}{\milli\m\squared}$ ($\Delta z \sim \SI{20}{\micro\m}$, $\Delta y \sim \SI{25}{\milli\m}$), the beam waist of the laser $\omega_0\sim\SI{0.75}{\milli\m}$ and the probability of the atoms with the given velocity distribution to have a velocity within the selected velocity interval, $P_v(v_{\rm{min}}\ge v_x \ge v_{\rm{max}})=3.9\times10^{-3}$. 
These parameters yield an estimated rate of $H$ passing through the \textsc{gqs} chamber of
\begin{multline}
     R=R_{\rm{in}} d P_v(v_{\rm{min}}\ge v_x \ge v_{\rm{max}}) \\
     \frac{A}{2\pi \Delta x^2\left(1-\cos{\theta/2}\right)}\frac{2\omega_0}{(v_{\rm{max}}-v_{\rm{min}})t_{\bar{v}}} 
      \cong \SI{E4}{\per\s}\, .
\end{multline}

Multiplying $R$ by the ionization efficiency of $\epsilon_{\rm{ion}}\sim 8\%$ (determined by $\omega_0$ and an assumed laser energy of \SI{1}{\milli\joule}) and the MCP efficiency $\epsilon_{\rm{MCP}}\sim50\%$ yields the signal rate $R_{\rm{sig}}\cong\SI{400}{\per\s}$.
This is $4\times10^3$ times more $H$ signal, as compared to the \textsc{ucn} signal.



\section{Outlook}
\label{sec:Outlook}
There  are currently three major improvements being implemented and tested.

New UV mirrors with a higher damage threshold were installed. It can be expected, that the laser intensity will be improved by an order of magnitude: when the beam size is compressed to $\omega_0\sim\SI{0.3}{\milli\m}$, with a UV laser energy of $\sim\SI{1}{\milli\J}$, the intensity becomes $\sim \SI{3.6E7}{\W\per\centi\m\squared}$. At this level, saturation is reached and the signal becomes independent of laser energy- or frequency fluctuations. Furthermore the ionization efficiency will be improved dramatically. A beam size of \SI{0.3}{\milli\m} yields an ionization efficiency of $\epsilon_{\rm{ion}}=98.22\%$, which would improve the estimated rate by a factor of 12.

The estimated rate will further be improved, by the installation of a new coldhead and an additional heatshield. This is currently being implemented, and first measurements show, that temperatures around $\sim\SI{4}{K}$ can be expected. This would improve our estimated rate for the velocity interval $[\SI{62}{\m\per\s},\ \SI{100}{\m\per\s}]$ by a factor of $\sim 2.2$. It would alternatively be possible to select slower velocities in the interval $[\SI{55}{\m\per\s},\ \SI{83}{\m\per\s}]$ while maintaining the same countrate as with the old cryo system.

It would be preferable to go to even lower velocities.
But, as can be seen in Fig. \ref{fig: delay} at around \SI{20}{\milli\s}, the residual hydrogen gas in the chamber prevents the measurements to be sensitive to atoms with lower velocities.
This could be improved by an aperture system between the source and the chamber where the \textsc{gqs} region will be installed. Such a system is currently being installed and tested.
It consists of three height adjustable, vertical slits with a width of \SI{200}{\micro\m} for the first slit and \SI{1}{\milli\m} for the second and third. Two more vacuum pumps will be installed in between the first and the second and the second and the third slit.
This system has two purposes: It will decrease the background, due to the separation of the different vacuum regions of the cryogenic chamber, the beamline and the detection chamber. It will also act as a velocity selecting aperture. As the slit height is adjustable, different trajectories of the atoms can be selected. With the three slits, it will be possible to select the low energy tail of the $H$ atoms with a vertical velocity component $v_z\sim0$ at the entrance of the \textsc{gqs} spectrometer as described in section \ref{sec:ucn_methods}.

As soon as those new implementations are completed and characterized, the \textsc{gqs} chamber will be installed at the end of the beamline replacing the detection chamber. It will contain the \textsc{gqs} spectrometer and the viewports for the UV laser. In this way, the atoms passing through the spectrometer will be photo ionized at the end of the mirror and the $H^+$ detected in the MCP.



\section{Conclusions}
\label{sec:conclusions}
We conclude that a \textsc{gqs} measurement with $H$ is a very promising but challenging endeavor. The expected count rate exceeds the count rates accessible with \textsc{ucn}s by orders of magnitudes. 

An extension and improvement of the existing \textsc{gqs} measurements is highly interesting for multiple fields. 

In the course of realizing a \textsc{gqs} measurement with $H$, we set up a cryogenic $H$-beam. A highly efficient $H$ detection system was developed. New UV mirrors, an improved cryogenic system and an aperture system which will reduce the background and select ideal velocity components are currently being implemented and tested.

We aim to demonstrate the existence of \textsc{gqs} of $H$ within the next measurement campaign, starting in 2023.

\section*{Acknowledgments}
This project was supported by the Austrian Science Fund (FWF) [W1252-N27] (Doktoratskolleg Particles and Interactions) and the ETH Zurich Career Seed Grant [SEED-17 20-1].
François Nez and Pauline Yzombard acknowledge support from CNRS (IEA 2021-2022 QRECH).
Paolo Crivelli acknowledges the support of the European Research Council (grant 818053-Mu-MASS) and the Swiss National Science Foundation (grant 197346).

\section*{Author contributions}
All authors contributed to the study conception and design. Material preparation, data collection and analysis were performed by CK, ZB, PB, OH, KS and PY. The first draft of the manuscript was written by CK and all authors commented on previous versions of the manuscript. All authors read and approved the final manuscript.

\subsection*{Data Availability Statement}
 The datasets generated during and/or analysed during the current study are available from the corresponding author on request.




\end{document}